# Field-free superconducting diode effect in noncentrosymmetric superconductor/ferromagnet multilayers

Hideki Narita[1]*, Jun Ishizuka[2], Ryo Kawarazaki[1], Daisuke Kan[1,3], Yoichi Shiota[1,3], Takahiro Moriyama[1,3], Yuichi Shimakawa[1,3], Alexey V. Ognev[4], Alexander S. Samardak[4], Youichi Yanase[5,6] and Teruo Ono[1,3,4,7]*

[1] Institute for Chemical Research, Kyoto University, Gokasho, Uji, Kyoto, 611-0011, Japan.

[2] Institute for Theoretical Physics, ETH Zurich, 8093 Zurich, Switzerland

[3] Center for Spintronics Research Network, Institute for Chemical Research, Kyoto University, Gokasho, Uji, Kyoto, 611-0011, Japan.

[4] Laboratory of Spin-Orbitronics, Institute of High Technologies and Advanced Materials, Far Eastern Federal University, Vladivostok 690950, Russia

[5] Department of Physics, Graduate School of Science, Kyoto University, Kyoto, 606-8502, Japan

[6] Institute for Molecular Science, Okazaki, 444-8585, Japan

[7] Center for Spintronics Research Network, Graduate School of Engineering Science, Osaka University, Machikaneyama 1-3, Toyonaka, Osaka 560-8531, Japan

* Correspondence: E-mail: narita.hideki.3x@kyoto-u.ac.jp (Hideki Narita); ono@scl.kyoto-u.ac.jp (Teruo Ono)

The diode effect is fundamental to electronic devices and is widely used in rectifiers and AC–DC converters. At low temperatures, however, conventional semiconductor diodes possess a high resistivity, which yields energy loss and heating during operation. The superconducting diode effect (SDE)[1-8], which relies on broken inversion symmetry in a superconductor may mitigate this obstacle: in one direction a zero-resistance supercurrent can flow through the diode, but for the opposite direction of current flow, the device enters the normal state with ohmic resistance. The application of a magnetic field can induce SDE in Nb/V/Ta superlattices with a polar structure[1,2], in superconducting devices with asymmetric patterning of pinning centres[9], or in superconductor/ferromagnet hybrid devices with induced vortices[10,11]. The need for an external magnetic field limits their practical application. Recently, a field-free SDE was observed in a $NbSe_2$/$Nb_3Br_8$/$NbSe_2$ junction, and it originates from asymmetric Josephson tunneling that is induced by the $Nb_3Br_8$ barrier and the associated $NbSe_2$/$Nb_3Br_8$ interfaces[12]. Here, we present another implementation of zero-field SDE using noncentrosymmetric $[Nb/V/Co/V/Ta]_{20}$ multilayers. The magnetic layers provide the necessary symmetry breaking and we can tune the SDE by adjusting the structural parameters, such as the constituent elements, film thickness, stacking order, and number of repetitions. We control the polarity of the SDE through the magnetization direction of the ferromagnetic layers. Artificially stacked structures[13-18], as the one used in this work, are of particular interest as they are compatible with microfabrication techniques and can be integrated with devices such as Josephson junctions[19-22]. Energy-loss-free SDEs as presented in this work may therefore enable novel non-volatile memories and logic circuits with ultralow power consumption.

In this regard, we fabricated noncentrosymmetric ferromagnet/superconductor multilayers [Nb (4.5 nm)/V (4.5 nm)/Co (1.7 nm)/V (4.5 nm)/Ta (4.5 nm)]$_{20}$ by replacing the V layers in Nb/V/Ta superlattices[1,2,23,24] with symmetric V/Co/V units to control the Rashba superconductor[25] by magnetization. The crystalline structure of the multilayers was characterized by low- and high-angle X-ray diffraction (XRD) profiles (see Methods and Supplementary Fig. S1). Figure 1a shows a schematic of the measurement configuration, wherein the magnetic field, electrical current, and polar axis are perpendicular to each other. Figure 1b shows a photomicrograph of the device with the polar axis along the *z*-direction. Electric current and magnetic field were applied along the *y*- and *x*-directions, respectively.

Figure 1c shows the temperature dependence of the resistivity of the device at an electrical current density of 0.455 kA/cm$^2$ under various magnetic fields ranging from 0 to 0.5 T.

The superconducting transition temperature, $T_c$ (the temperature at which the resistivity drops to half of that above the transition), of the device was determined to be 2.6 K in the absence of a magnetic field. The $T_c$ of the present device is lower than that of the previously studied [Nb (1nm)/V (1 nm)/Ta (1 nm)]$_{40}$ device[1]. This is possibly due to the pair-breaking effect of the Co layers[16] and/or the difference in the repetition number (see Supplementary Fig. S2).

Furthermore, vortices were observed to start breaking the superconducting phase with increasing magnetic field, and the superconducting phase completely disappeared above 0.5 T at the lowest measurement temperature of 1.9 K. The superconducting coherence length ξ(0) was estimated to be 24 nm (see Supplementary Fig. S3). The magnetization curves of the multilayers at different temperatures in Fig. 1d show the characteristic ferromagnetic responses with a coercive field of 0.065 T in the multilayers above and below the $T_c$.

Figure 1e presents the magnetic field dependences of the resistivity at several electrical current densities evaluated at 1.9 K. We observed switching between normal and superconducting states, similar to that observed in previous studies[13,14]. This is caused by the interplay between superconductivity and ferromagnetism. Previous reports have suggested that the switching between the normal and superconducting states is due to the internal exchange field at the superconductor/ferromagnet interface[13,14]. In our experiments, the switching between the normal and superconducting states was observed in the current density range of 0.455–20.45 kA/cm$^2$. Because of the suppression of superconductivity by the external magnetic fields, the electrical resistivity increases monotonically with the magnetic fields at a low current density of 0.455 kA/cm$^2$.

As the current density increases, the hysteresis of magnetoresistance becomes more pronounced, and the conventional magnetoresistance of Co is observed above 22.72 kA/cm$^2$, where the superconductivity is completely destroyed. The observation of the conventional magnetoresistance of Co also indicates that the device fabrication process does not alter the ferromagnetic properties of the multilayers.

Figures 2a and 2b show the resistivity as a function of the current density, $J$, under various magnetic fields for positive and negative currents at 1.9 K. We observed the SDE in our noncentrosymmetric superconductor/ferromagnet multilayers; the critical current density $J_c$ (the current density at which the resistivity increases to half of that above the transition) depends on the polarity of the current. Note that $J_c$ is determined by the transition from the superconducting state to the normal conductive state. Therefore, there is no influence from the Joule heating. The variation in $J_c$ with the magnetic field at 1.9

K is shown in Fig. 2c. In these $J_c$ measurements, the magnetic field was initially set to +0.5 T, which is sufficiently large to saturate the magnetization, determined from the results in Fig. 1c. Subsequently, $J_c$ was measured by sweeping the magnetic field in the order of +0.5, 0, −0.5, 0, +0.5 T. $J_c$ also exhibits hysteresis, reflecting the magnetic hysteresis of the Co layer. $J_c$ shows a symmetric behavior with respect to zero field (dashed line) for the sweep of the magnetic field in the order of −0.5, 0, +0.5, 0 T. The pronounced peak around the coercive field can be attributed to enhanced vortex pinning on the network of proliferating domain walls[26].

Figure 2d shows the magnetic field dependences of the nonreciprocal components of the critical current densities, $\Delta J_c = |+J_c| - |-J_c|$, at different temperatures ($+J_c$ and $-J_c$ are the critical current densities corresponding to positive and negative currents, respectively). Here, the magnetic field was initially set to +0.5 T. Then, $\Delta J_c$ was measured by sweeping the magnetic field in the order of +0.5, 0, −0.5, 0, +0.5 T. In the downward sweep from the magnetic field of +0.5 T at 1.9 K (black dots), a negative peak of $\Delta J_c$ appears at approximately +0.008 T, and $\Delta J_c$ approaches zero at 0 T. As the magnetic field decreases further, large negative and positive peaks of $\Delta J_c$ appear at approximately −0.008 and −0.04 T, respectively. $\Delta J_c$ then gradually decreases to zero at around −0.5 T. After saturating the magnetization at −0.5 T, $\Delta J_c$ was measured again by sweeping the magnetic field in the order of −0.5, 0, +0.5 T. Moreover, the magnetic field dependence of $\Delta J_c$ in the upward sweep shows a trend (red dots) opposite to that of the downward sweep (black dots). The microscopic origin of the observed peaks and dip structures in the magnetic field dependence of $\Delta J_c$ is not clear. However, it is possibly related to the evolution of complex magnetic domains. We also checked the reproducibilities of $J_c$ and $\Delta J_c$ in additional two devices (see Supplementary Fig. S4), and we observed that there was no clear hysteresis of $\Delta J_c$ in the configurations of **B** // **x** and **B** // **z** (see Supplementary Fig. S5). This hysteresis behavior of $\Delta J_c$ for **B** // **y** is in clear contrast with the SDE observed in Nb/V/Ta superlattices[1], wherein no hysteresis is observed. Note that the symmetric [Nb/V/Co/V/Nb]$_{20}$ multilayers without a polar structure do not exhibit the clear nonreciprocal components of the critical current densities even though they exhibit hysteresis in the magnetization curves (see Supplementary Fig. S6). Therefore, the polar structure is essential to realize the magnetization-mediated SDE.

The polarity control of SDE in a zero field is essential to develop non-volatile devices based on SDEs. We demonstrate it using the minor hysteresis loops of the magnetization. Figure 3a shows the magnetic field dependences of $J_c$ for positive and negative currents in minor hysteresis loops. For this, the magnetic field was swept in the order of +0.5, 0, −0.15 T (downward sweep) and then −0.15, 0, +0.15 T (upward sweep). The limiting field of the minor loops is 0.15 T, as shown by the dashed line in Fig. 1d. Owing to the asymmetrical magnetic field sweeping, $J_c$ shows asymmetry with respect to the zero field (dashed line). Consequently, we observe the positive $\Delta J_c$ in the zero-field after sweeping the magnetic field in the order of +0.5, 0, −0.15, 0 T, as shown in Fig. 3b. Moreover, $J_c$ also shows asymmetry with respect to zero field after sweeping the magnetic field in the order of −0.5, 0, +0.15, 0 T, resulting in a negative $\Delta J_c$ under the zero-field condition, as shown in Figs. 3c and 3d, respectively. Therefore, we could control the polarity of SDE by magnetization.

Here, we demonstrate the non-volatile SDE. The red and black dots in Fig. 3e represent the resistivities measured with $J = 72.7$ kA/cm$^2$ at 1.9 K for negative (−**M**) and positive magnetizations (+**M**), respectively. The −**M** (+**M**) is the state after sweeping the magnetic field in the order of +0.5, 0, −0.15, 0 T (−0.5, 0, +0.15, 0 T). The device exhibits a superconducting state or a normal conducting state depending on the polarity of the current, and the polarity of SDE depends on the direction of magnetization.

Finally, we discuss theoretical viewpoints in detail. We performed first-principles band structure calculation in the nonmagnetic electronic state for a [Nb/V/Co/V/Ta] supercell, where two atomic layers of bcc-Nb, V, Co, V, and Ta were repeatedly stacked, as shown in Figs. 4a and b. The Rashba splitting of bands at the Fermi energy $E_F$ was observed as $E_R = 1–10$ meV, which is close to the value of a [Nb/V/Ta] superlattice. Although the Rashba splitting of Co bands was minor (~ 1 meV), the density of states due to the Co bands near $E_F$ was predominantly larger than that due to the Nb, V, and Ta bands, as shown in Fig. 4b. To gain more insight, we also performed a spin-polarized band structure calculation with the magnetization direction (010) and obtained a ferromagnetic ground state, where the total energy was lower than the nonmagnetic state, as shown in Figs. 4c and d. The magnitude of magnetic moment $|M|$ at Co atoms was calculated as $|M| = 1.93$ $\mu_B$. The V atoms adjacent to Co atoms had a considerable magnetization $|M| = 0.5–1$ $\mu_B$, while others had only $|M| = 0.01–0.05$ $\mu_B$. Thus, the atoms that are not adjacent to Co atoms can stabilize the superconductivity against the Pauli depairing effect due to the ferromagnetic spin polarization. We indeed noticed that the partial densities of states are almost unchanged for Nb, V, and Ta atoms, and all atoms contribute at $E_F$ in the ferromagnetic state, as shown in Fig. 4d. In other words, the superconducting transition temperature of this system may be determined from the total density of states consisting of three types of atoms.

Pairing states that exhibit the diode effect would correspond to helical superconductivity, which has been investigated in noncentrosymmetric superconductors such as CePt$_3$Si[27,28], the surface/interface of SrTiO$_3$/LaAlO$_3$[29], and heavy-fermion superlattices[30]. The asymmetric band dispersion is essential for the helical superconductivity, and it appears in the noncentrosymmetric ferromagnet. Our spin-polarized band structure calculation indeed shows asymmetric behaviors (see Supplementary Fig. S7). The SDE has been analyzed by the Ginzburg-Landau (GL) theory and microscopic Hamiltonian[3, 6-8]. Higher-order gradient terms in the GL expansion exhibit the nonreciprocity of critical current, and the microscopic calculation predicts an anomalous "sign-reversal effect" in the helical superconducting state[6]. Thus, it is intriguing to clarify the asymmetric and sign-reversing behavior observed in Fig. 2d in future works.

The experiments revealed that noncentrosymmetric superconductor/ferromagnet multilayers exhibited magnetization-mediated nonreciprocal critical currents, *i.e.*, a magnetization-mediated SDE. The field-free SDE demonstrated here enables rectification without an external magnetic field. However, significant issues still need to be overcome for practical application. The rectification still requires precise controls of the applied current and temperature, which are major limitations for practical device operation. Therefore, it is essential to explore materials with a larger $\Delta J_c$ and higher $T_c$ to realize a wide temperature range of operation. The proposed method of symmetry breaking using artificial stacked structures is versatile and controllable. Furthermore, the electronic band structure can be controlled by modulating the thickness and period of each constituent layer, which offers a promising route for exploring new materials for practical application. Our findings pave the way for obtaining rewritable superconducting diode logic circuits using the nonvolatility of magnetic materials.

**Acknowledgments**

This work was partly supported by JSPS KAKENHI (Grant No. 15H05702, 15H05884, 15H05745, 18H04225, 18H01178, 18H05227, 19H05823, 20H05665, 21K13883 and 21K18145); the Cooperative Research Project Program of the Research Institute of Electrical Communication, Tohoku University; and the Collaborative Research Program of the Institute for Chemical Research, Kyoto University. This study was also supported by the Futaba Foundation's Futaba Research Grant Program, and Iketani Science and Technology Foundation. A. V. O, A. S. S. and T. O. acknowledge the Russian Ministry of Science and Higher Education for state support of scientific research conducted under the supervision of leading scientists in Russian institutions of higher education, scientific foundations and state research centers (Project No. 075-15-2021-607).

**Author contributions**

H. N. and T. O. conceived the project. H. N. deposited the multilayers and fabricated the devices. H. N. performed the measurements and data analysis. H. N. and T. O. wrote the manuscript with the assistance of the other authors. J. I. and Y. Y. calculated the band structure and assisted with the analysis of the experimental results. All authors contributed jointly to the interpretation of the results.

**Competing Interests**

The authors declare no competing financial interests.

**Figure Legends**

**Fig. 1. Device structure, transport, magnetic properties, and SDE. a**, Schematic of the SDE and measurement configuration. The magnetic field is applied perpendicular to both the polar axis and the electrical current. **b**, Photomicrograph of the device. The wires of the [Nb/V/Co/V/Ta]$_{20}$ multilayers were electrically connected to the Au (100 nm)/Ti (5.0 nm) metal electrodes. **c**, Temperature dependence of the device resistivity at an electrical current density (0.455 kA/cm$^2$) in the multilayers under a magnetic field in the range of 0 to 0.5 T. **d**, Magnetization loops at several temperatures for the multilayers above and below $T_c$. **e**, Magnetic field dependence of the resistivity in the multilayers for several electrical current densities at 1.9 K. The direction of the magnetic field was perpendicular to the polar axis and the electrical current.

**Fig. 2. Magnetic field and temperature dependences of SDE. a**, Current density ($J$) dependence of the resistivity under various magnetic fields for positive and negative currents at 1.9 K. The magnetic field was swept from +0.5 T to −0.5 T (downward sweep). **b**, Dependence of the resistivity on $J$ under various magnetic fields for positive and negative currents at 1.9 K. The magnetic field was swept from −0.5 T to +0.5 T (upward sweep). **c**, $J_c$ as a function of the magnetic field at 1.9 K. First, the magnetic hysteresis of $J_c$ was measured from +0.5 to −0.5 T (downward sweep) for positive (red) and negative (black) currents. Subsequently, the sweep direction was reversed, and the magnetic field was swept from −0.5 to +0.5 T (upward sweep) for positive (blue) and negative (green) currents. The arrows indicate the sweep directions. **d**, Nonreciprocal component of the critical current density, $\Delta J_c$, as a function of the magnetic field at various temperatures. The magnetic hysteresis of $\Delta J_c$ was measured from +0.5 to −0.5 T (downward sweep, black). Subsequently, the sweep direction was reversed, and the magnetic field was swept from −0.5 to +0.5 T (upward sweep, red). The arrows indicate the sweep direction.

**Fig. 3. Field-free SDE controlled by magnetization**  **a**, Magnetic field dependences of $J_c$ for positive and negative currents in minor hysteresis loops at 1.9 K. Magnetic field was swept in the order of +0.5, 0, −0.15 T (downward sweep) and then −0.15, 0, +0.15 T (upward sweep). **b**, Magnetic field dependence of $\Delta J_c$ in the minor hysteresis loops obtained by Fig. 3**a**. **c**, Magnetic field dependences of $J_c$ for positive and negative currents in minor hysteresis loops at 1.9 K. Magnetic field was swept in the order of −0.5, 0, +0.15 T (upward sweep), and then +0.15, 0, −0.15 T (downward sweep). **d**, Magnetic field dependence of $\Delta J_c$ in the minor hysteresis loops obtained by Fig. 3**c**. **e**, Repeated application of current densities $J = 72.7$ kA/cm$^2$ and $J = -72.7$ kA/cm$^2$ at 1.9 K without a magnetic field. **f**, Non-volatile SDE at 1.9 K. Red and black dots represent the results for negative magnetization (-*M*) and positive magnetization (+*M*), respectively. The device shows a superconducting state or normal conducting state depending on the polarity of the current. Note that the polarity of SDE depends on the direction of magnetization. The -*M* or +*M* state is achieved after sweeping the magnetic field in the order of +0.5, 0, −0.15, 0 T or −0.5, 0, +0.15, 0 T.

**Fig. 4. Band structure of a bulk [Nb/V/Co/V/Ta] superlattice**
**a**, Band structure obtained from first-principles calculation for paramagnetic state. **b**, Partial density of states for paramagnetic state. **c**, Band structure for ferromagnetic state. **d**, Partial density of states for ferromagnetic state.

**METHODS**

**Device fabrication**

The [Nb/V/Co/V/Ta]$_{20}$ multilayers were grown on MgO (100) substrates by direct current (DC) magnetron sputtering in a high-vacuum system with a base pressure of approximately $5 \times 10^{-6}$ Pa. Before the film growth, the MgO (100) substrate was heated at 600 °C for 30 min in a sputtering chamber to remove impurities. Subsequently, the MgO substrate was heated to 300 °C during the deposition. All the films were deposited under an Ar pressure of approximately 0.1 Pa. The Nb, V, Co, V, and Ta layers were repeatedly sputtered on the MgO substrate in the same sequence for 20 cycles. The deposition rates were maintained constant at 0.36, 0.22, 0.45, and 0.18 Å/s for Nb, V, Ta, and Co, respectively. The total deposition time was approximately 5 h. After deposition, the films were then cooled from 300 to 30 °C. XRD was performed using a conventional 4-circle diffractometer (PANalytical, X'pert MRD) with Cu-K$_{\alpha1}$ radiation ($\lambda$ = 0.154 nm). Supplementary Fig. S1a and b show the low- and high-angle XRD profiles of the multilayers, respectively. In this study, the thickness of the ferromagnetic layer was adjusted to realize ferromagnetic properties, and it was set to be sufficiently small to avoid the Bloch domain wall effect.

The deposited films were then patterned into a wire (dimensions: 50 μm × 1000 μm × 0.44 μm) using conventional photolithography and Ar-ion milling. Finally, Au (100 nm)/Ti (5.0 nm) metal electrodes were electrically connected to the wire. The surface layer was removed through weak Ar-ion milling before the electrode deposition to create Ohmic contacts.

**Transport and magnetization measurements**

During the transport measurements, the temperature and magnetic field were controlled using a commercial refrigerator (Quantum Design, Physical Property Measurement System) with DC applied using a current source (Yokogawa 7651). The voltage was measured using a four-terminal configuration with a nanovoltmeter (Keithley 2182A).

The magnetization of the sample was studied under a magnetic field parallel to the film surface, using a superconducting quantum interference device magnetometer (Quantum Design MPMS3). Before starting the measurements, the magnetic field was reduced from 1.0 to 0 T using an oscillating field sequence to remove any trapped flux in the superconducting magnet. Subsequently, the magnetic hysteresis loops of the zero-field-cooled magnetization were measured in the "DC and no overshoot" mode using film samples with dimensions of 7 mm × 2 mm. The magnetization of the MgO substrate was subtracted to determine the intrinsic magnetic properties of the multilayers with the field parallel to the surface.

**Details of band structure calculation**

We conducted first-principles calculations for nonmagnetic and ferromagnetic states on a bulk [Nb/V/Co/V/Ta] superlattice. We adopted the full-potential linearized augmented plane wave+local orbitals method within the generalized gradient approximation in the WIEN2k package[31, 32]. A supercell was constructed with ten layers consisting of five two-layers of Nb, V, Co, V, and Ta atoms. We used the muffin-tin radius $R_{MT}$ of 2.50 a.u., the maximum reciprocal lattice vector $K_{max}$ as $R_{MT}K_{max} = 8.0$, and $21 \times 21 \times 4$ $k$-points sampling for self-consistent calculation. Figure S7 shows the band structures and the partial densities of states in nonmagnetic and ferromagnetic states. If we neglect the Co bands, the band structure is similar to that in the [Nb/V/Ta] superlattice. The partial densities of the states of Co show an exchange splitting, while they are almost unchanged for Nb, V, and Ta. Thus, Nb, V, and Ta layers form spin-splitting bands with weak magnetization. Therefore, it is verified that [Nb/V/Co/V/Ta] is a noncentrosymmetric superconductor that exhibits a SDE without a magnetic field.

**Data Availabillity**

The data that support the findings of this study are available from the corresponding authors upon reasonable request.

**Methods-only references**

Fig. 1

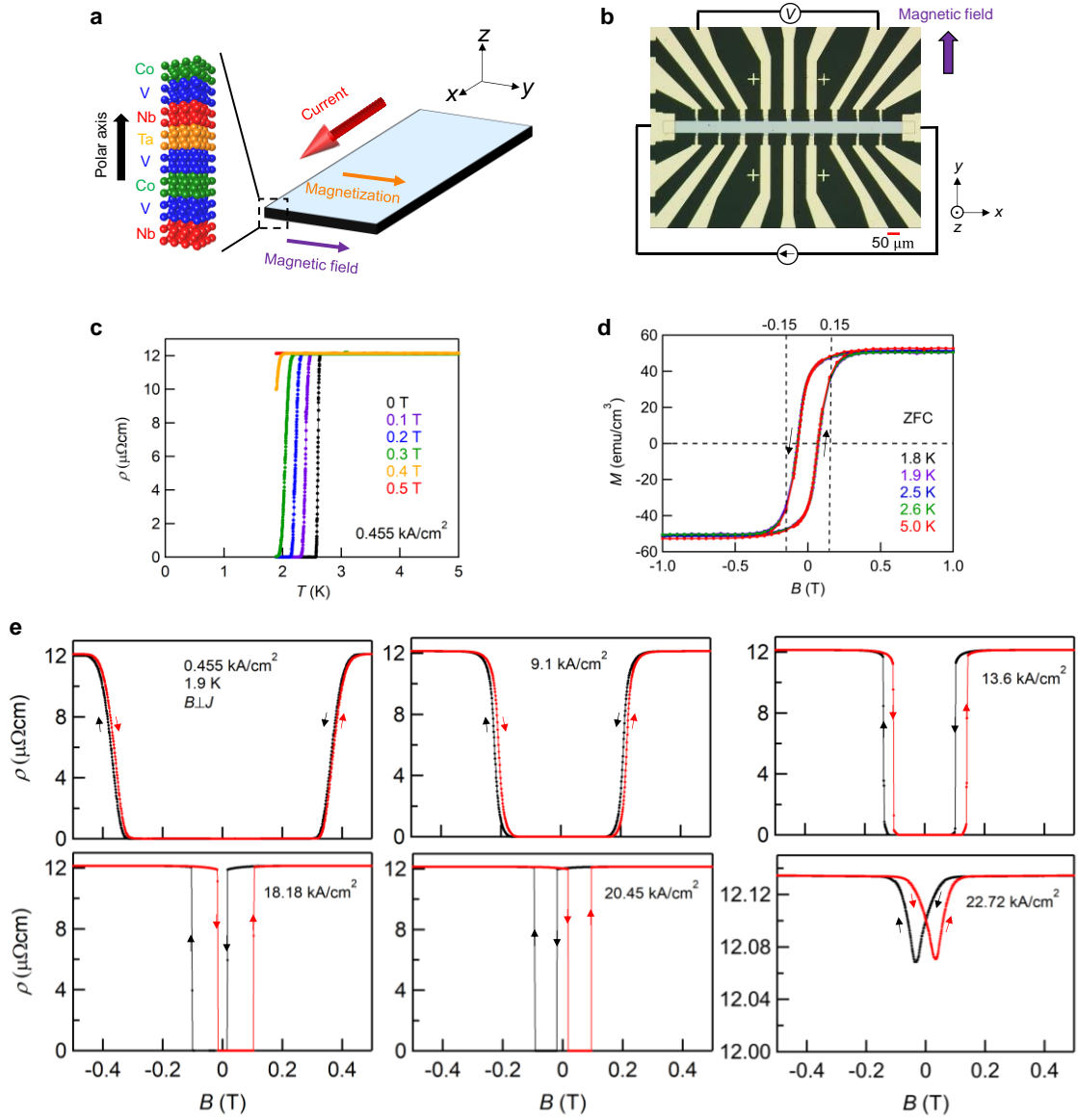

Fig. 2

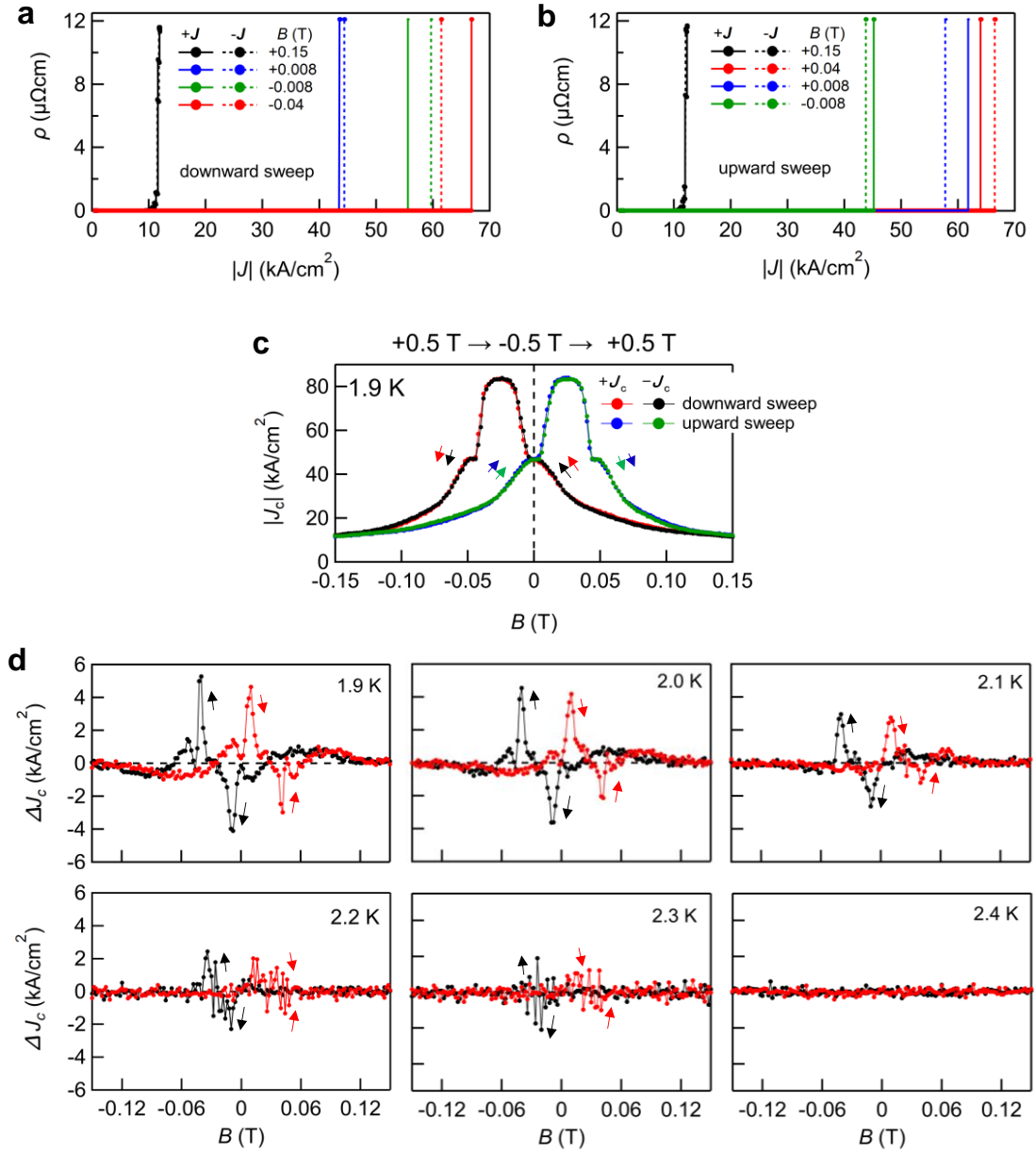

Fig. 3

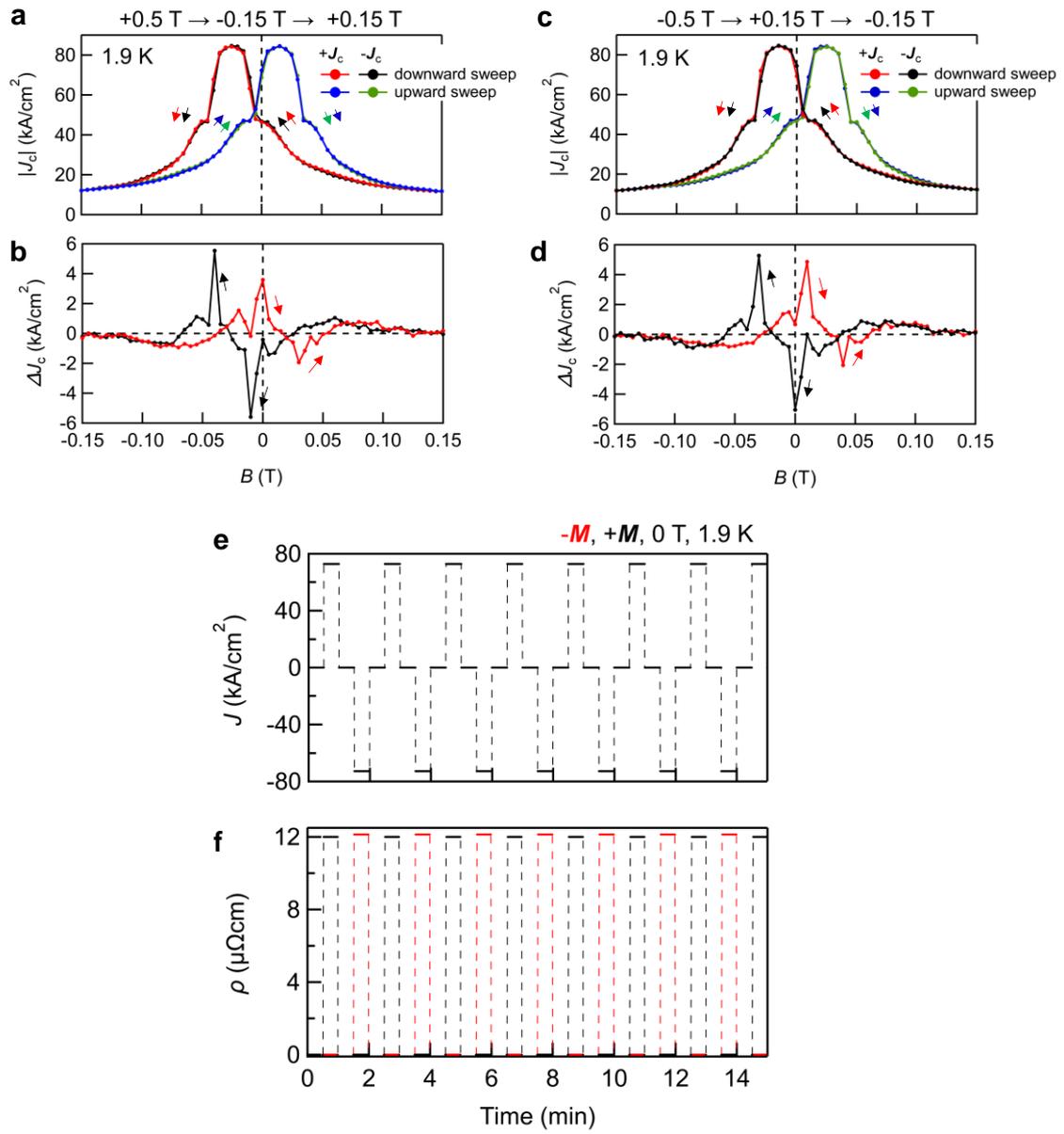

Fig. 4

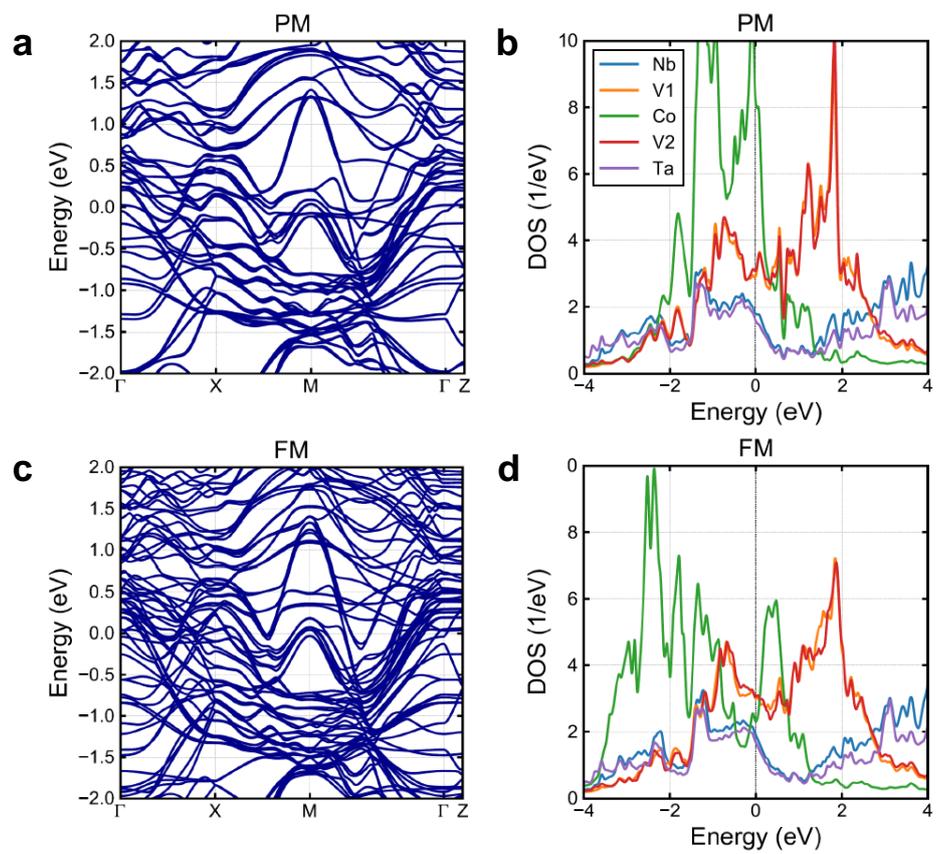

# Supplementary information

# Field-free superconducting diode effect in noncentrosymmetric superconductor/ferromagnet multilayers

Hideki Narita[1]*, Jun Ishizuka[2], Ryo Kawarazaki[1], Daisuke Kan[1,3],

Yoichi Shiota[1,3], Takahiro Moriyama[1,3], Yuichi Shimakawa[1,3], Alexey V.

Ognev[4], Alexander S. Samardak[4], Youichi Yanase[5,6] and Teruo Ono[1,3,4,7]*

[1] Institute for Chemical Research, Kyoto University, Gokasho, Uji, Kyoto, 611-0011, Japan.

[2] Institute for Theoretical Physics, ETH Zurich, 8093 Zurich, Switzerland

[3] Center for Spintronics Research Network, Institute for Chemical Research, Kyoto University, Gokasho, Uji, Kyoto, 611-0011, Japan.

[4] Laboratory of Spin-Orbitronics, Institute of High Technologies and Advanced Materials, Far Eastern Federal University, Vladivostok 690950, Russia

[5] Department of Physics, Graduate School of Science, Kyoto University, Kyoto, 606-8502, Japan

[6] Institute for Molecular Science, Okazaki, 444-8585, Japan

[7] Center for Spintronics Research Network, Graduate School of Engineering Science, Osaka University, Machikaneyama 1-3, Toyonaka, Osaka 560-8531, Japan

* Correspondence: E-mail: narita.hideki.3x@kyoto-u.ac.jp (Hideki Narita); ono@scl.kyoto-u.ac.jp (Teruo Ono)

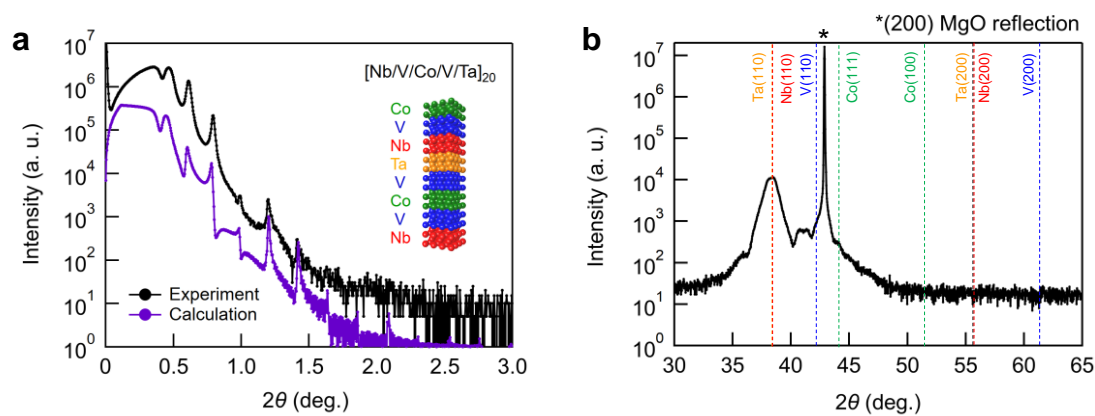

**Fig. S1. Low- and high-angle XRD profiles. a**, Measured (black) and simulated (purple) low-angle XRD profiles of the [Nb/V/Co/V/Ta]$_{20}$ multilayer. **b**, High-angle XRD profile of the multilayer.

Figure S1a shows the design of the [Nb/V/Co/V/Ta]$_{20}$ multilayer structure and its low-angle X-ray diffraction (XRD) profile. The measured curve was fitted as a function of the thickness and roughness of the Nb, V, Ta, and Co layers using the PANalytical X'Pert reflectivity program. Furthermore, the simulated curve reproduced the measured Bragg and Kiessing fringes. The calculated thicknesses of the Nb, V, Ta, and Co layers were approximately 4.5, 4.5, 4.5, and 1.7 nm, respectively, which did not precisely match the original design parameters of the multilayers [Nb (5 nm)/V (5 nm)/Co (2 nm)/V (5 nm)/Ta (5 nm)]$_{20}$, because the mean value of the interfacial roughnesses obtained from the fit is approximately 1.0 nm. Co layers with a thickness of 1.7 nm can serve as pair-breaking layers, thus suppressing superconductivity and completely decoupling the Nb layers[1].

Supplementary Fig. S1b shows the high-angle XRD profile around the (200) MgO reflection for the multilayers. The dotted lines in the figure represent the $2\theta$ positions of the reflections expected from the constituent layers of the multilayer structure.

The peak at $2\theta \approx 38°$ corresponds to the overlapped (110) peaks expected from bcc-Nb and bcc-Ta. However, distinguishing the expected peaks of the bcc-V and Co layers is difficult because of their overlap with the (200) MgO peak. In addition, the (200) peaks of the bcc-Nb, Ta, and V layers are not observed. Therefore, the fabricated multilayers mainly consist of layers with a bcc structure and are preferentially oriented along the (110) direction. The grain size was estimated to be 15 nm from the full-width at half-maximum of the (110) peak using the characteristic wavelength of the Cu-K$_{\alpha 1}$ radiation, full-width at half-maximum = 0.0118 rad, $\theta = 0.669$ rad, and Scherrer constant $K = 0.89$.

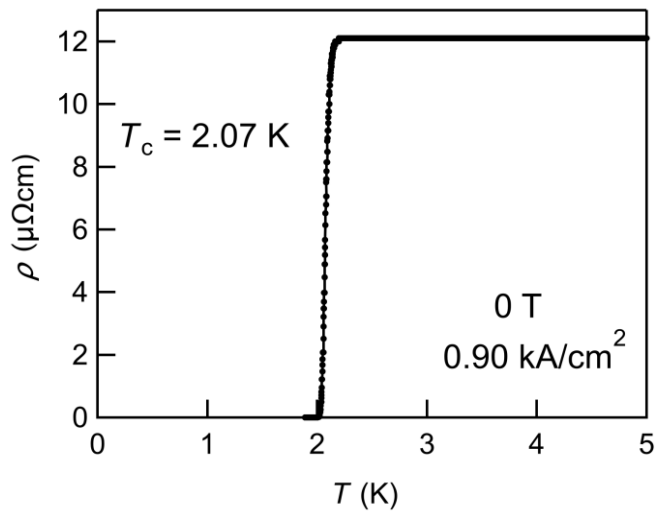

**Fig. S2 Transport properties in [Nb (4.5 nm)/V (4.5 nm)/Co (1.7 nm)/V (4.5 nm)/Ta (4.5 nm)]$_{10}$ multilayers.** Dependence of the resistivity of the device on the temperature with an electrical current density of 0.90 kA/cm$^2$ in the multilayers in the absence of a magnetic field.

The orbital limiting field $B_{c2}^{orb}$ (the critical magnetic field at which the resistivity drops to half the value obtained above the transition) is given as $B_{c2}^{orb}(T) = \Phi_0/2\pi\xi^2(T)$, where $\Phi_0 = 2.07 \times 10^{-15}$ [T m$^2$] is the flux quantum. The superconducting coherence length was estimated from the temperature dependence of $B_{c2}$. Note that, $B_{c2}^{orb}(0)$ is usually calculated from the initial slope of the plot of $B_{c2}$ versus temperature around the $T_c$ using the Werthamer–Helfand–Hohenberg formula, which is expressed as, $B_{c2}^{orb}(0) = -0.69 T_c (dB_{c2}/dT)_{T_c}$ in the dirty limit[2]. As a result of linear fitting, we evaluated $B_{c2}^{orb}(0)$ (~ 0.58 T) and the superconducting coherence length $\xi(0)$ (~ 24 nm).

The Pauli limit in a BCS superconductor can be estimated using the following equation: $B_{c2}^{Pauli}(0) \sim 1.84\, T_c$ [T]. We evaluated $B_{c2}^{Pauli}(0)$ (~ 4.8 T) using $T_c$ (~ 2.6 K) in the [Nb (4.5 nm)/V (4.5 nm)/Co (1.7 nm)/V (4.5 nm)/Ta (4.5 nm)]$_{20}$ multilayers. Therefore, the out-of-plane upper critical field is determined by the orbital pair-breaking effect rather than the Pauli pair-breaking effect.

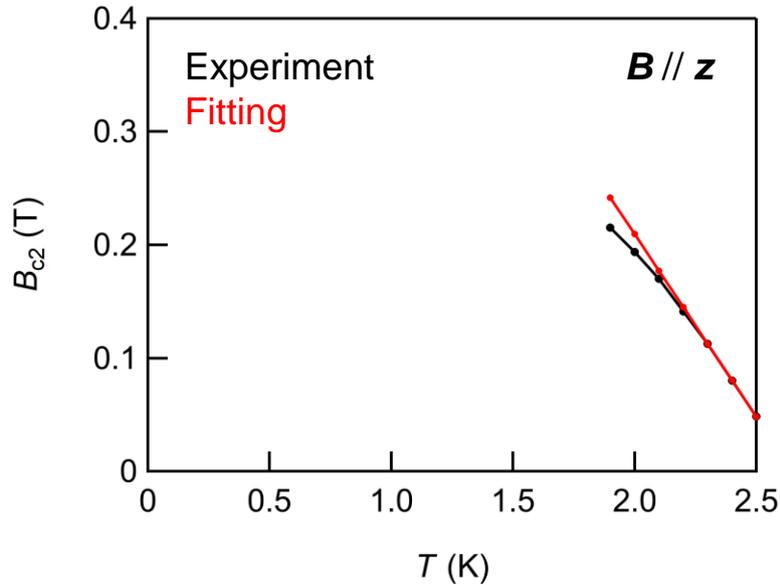

**Fig. S3 Temperature dependence of upper critical field $B_{c2}$**

$B_{c2}$ as a function of temperature at $J = 0.455$ kA/cm$^2$.

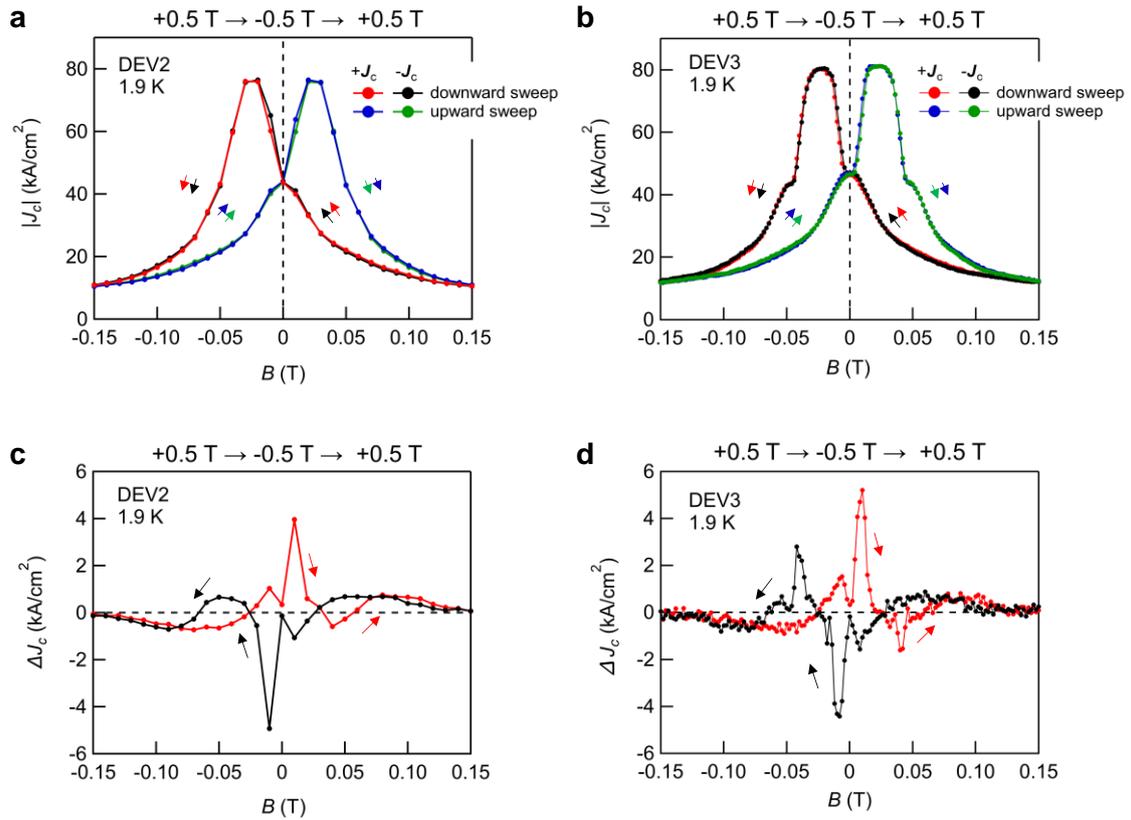

**Fig. S4 Reproducibilities of transport properties**

**a**, **b**, $J_c$ as a function of the magnetic field along the ***y*** axis at 1.9 K for other devices (DEV2 and DEV3) composed of asymmetric ferromagnet/superconductor [Nb (4.5 nm)/V (4.5 nm)/Co (1.7 nm)/V (4.5 nm)/Ta (4.5 nm)]$_{20}$ multilayers. **c**, **d**, Nonreciprocal component of the critical current density, $\Delta J_c$, as a function of the magnetic field at 1.9 K for DEV2 and DEV3.

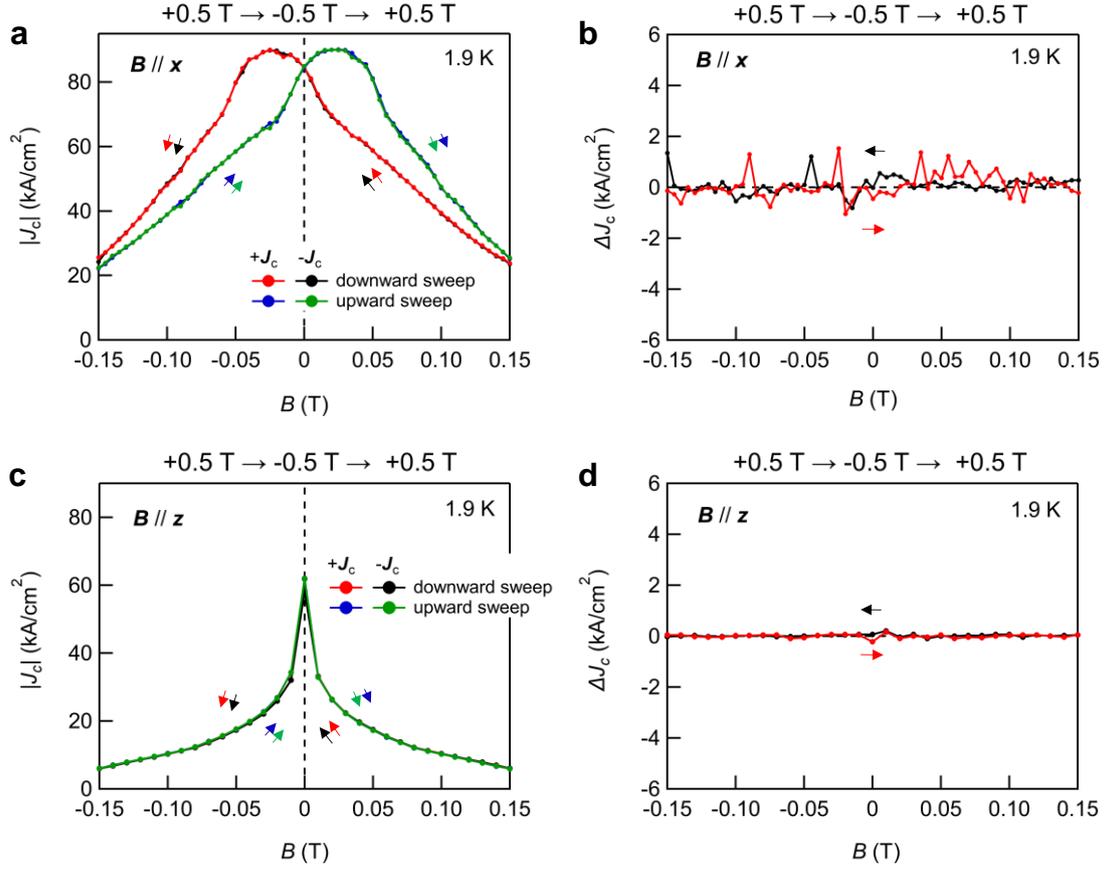

**Fig. S5 Transport properties in the configurations of $B \parallel x$ and $B \parallel z$.**

**a**, $J_c$ as a function of the magnetic field along the $x$-axis at 1.9 K. **b**, Nonreciprocal component of the critical current density, $\Delta J_c$, as a function of magnetic field at 1.9 K for the asymmetric ferromagnet/superconductor [Nb (4.5 nm)/V (4.5 nm)/Co (1.7 nm)/V (4.5 nm)/Ta (4.5 nm)]$_{20}$ multilayers. **c** $J_c$ as a function of the magnetic field along the $z$-axis at 1.9 K. **d** Nonreciprocal component of the critical current density, $\Delta J_c$, as a function of magnetic field at 1.9 K for the asymmetric ferromagnet/superconductor [Nb (4.5 nm)/V (4.5 nm)/Co (1.7 nm)/V (4.5 nm)/Ta (4.5 nm)]$_{20}$ multilayers.

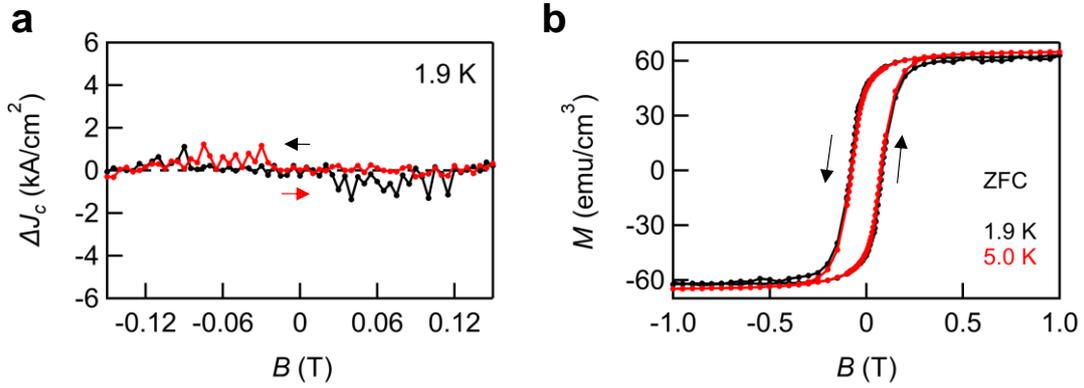

**Fig. S6 Transport and magnetic properties in [Nb (4.5 nm)/V (4.5 nm)/Co (1.7 nm)/V (4.5 nm)/Nb (4.5 nm)]$_{20}$ multilayers.**

**a**, Nonreciprocal component of the critical current density, $\Delta J_c$, as a function of magnetic field at 1.9 K in the symmetric ferromagnet/superconductor [Nb (4.5 nm)/V (4.5 nm)/Co (1.7 nm)/V (4.5 nm)/Nb (4.5 nm)]$_{20}$ multilayers. **b**, Magnetization loops for the multilayers above and below the $T_c$.

The [Nb (4.5 nm)/V (4.5 nm)/Co (1.7 nm)/V (4.5 nm)/Nb (4.5 nm)] multilayers, with $T_c$ = 3.74 K, were grown on MgO (100) substrates using the same method as that used for depositing [Nb (4.5 nm)/V (4.5 nm)/Co (1.7 nm)/V (4.5 nm)/Ta (4.5 nm)]. The crystalline structure of the multilayers was characterized using XRD measurements. To measure $\Delta J_c$, the multilayers were used to fabricate a device with the same dimensions as those shown in Fig. 1b in the main text. For the $\Delta J_c$ measurement, the magnetic field was initially set to +0.5 T. Then, $\Delta J_c$ was measured by sweeping the magnetic field in the order of +0.5, 0, −0.5, 0, and +0.5 T.

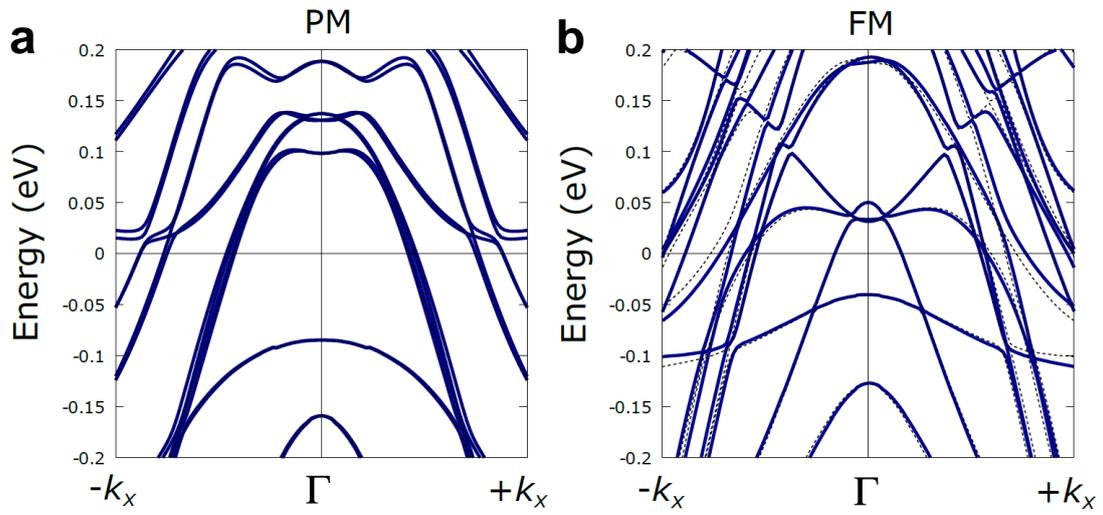

**Fig. S7. Low-energy electron band structure of a bulk [Nb/V/Co/V/Ta] superlattice.**

**a**, Low-energy electron band near the $\Gamma$ point for paramagnetic state. **b**, Low-energy electron band near the $\Gamma$ point for ferromagnetic state. The energy dispersion $E(-k_x, 0, 0)$ is shown in the dashed line as a guide for eyes.